\begin{filecontents*}{example.eps}
\documentclass[epjST]{svjour}
\usepackage{latexsym}
\usepackage{graphicx}
\usepackage{amssymb}
\begin{document}

\title{Analytical study of chaos and applications}
\author{George Contopoulos, Mirella Harsoula and Christos Efthymiopoulos}
\institute{Research Center for Astronomy and Applied Mathematics of the Academy of Athens\\
               \email{gcontop@academyofathens.gr}}

\date{Received: date / Revised version: date}

\abstract{
We summarize various cases where chaotic orbits can be described analytically.
 First we consider the case of a magnetic bottle where we have non-resonant and resonant ordered and chaotic orbits.
 In the sequence we consider the hyperbolic H\'enon map, where chaos appears mainly around the origin, which is an unstable periodic orbit.
 In this case the chaotic orbits around the origin are represented by analytic series (Moser series). We find the domain of convergence
  of these Moser series and of similar series around other unstable periodic orbits. The asymptotic manifolds from the various unstable periodic
   orbits intersect at homoclinic and heteroclinic orbits that are given analytically. Then we consider some Hamiltonian systems and we find
   their homoclinic orbits by using a new method of analytic prolongation. An application of astronomical interest is the domain
   of convergence of the analytical series that determine the spiral structure of barred-spiral galaxies.}

\maketitle

\section{Introduction}
\label{intro}

In generic nonintegrable dynamical systems there are both ordered and chaotic orbits.
In particular ordered orbits appear near stable periodic orbits, while chaotic orbits appear near unstable periodic orbits (for a review see Contopoulos 2002).

Let us consider a system of two degrees of freedom with Hamiltonian

\begin{equation}
 H=H(\rho,z,\dot{\rho},\dot{z})=H_{2}+H_{3}+\dots
\end{equation}

\noindent where

\begin{equation}
H_{2}=\frac{1}{2}({\dot{\rho}}^{2}+\dot{z}^{2}+{\omega_1}^{2} \rho^{2}+{\omega_2}^{2} z^{2})
\end{equation}

\noindent and $H_{s}$ contains terms of degree $s$.

Then the energy $H=E$ is an integral of motion around an equilibrium point $\rho=z=\dot{\rho}=\dot{z}=0$. When the equilibrium is stable we can find another integral of motion around it

\begin{equation}
 \Phi=\Phi(\rho,z,\dot{\rho},\dot{z})
\end{equation}

\noindent but this is in general only formal. If we develop $\Phi$ in power series around the origin

\begin{equation}
 \Phi=\Phi_{2}+\Phi_{3}+...
\end{equation}

\noindent  and truncate this series at an order $s$ the truncated
integral, $\overline{\Phi}_{s}$, is better conserved as the order of
truncation increases but only up to an optional order $s_{opt}$.
However, if we include also higher order terms the approximation
becomes worse. In fact the formal series $\Phi$ is divergent. Its
divergence is due to the appearance of small divisors of the form
$m_1\omega_1+m_2\omega_2$ with $m_1~,m_2$ integers.

We can now solve the system $(H, \Phi_{s})$ for a particular value of $\rho$ (e.g. $\rho=0$) and find a series

\begin{equation}
 f(z,\dot{z})=f_{2}+f_{3}+...
\end{equation}

\noindent that gives the successive points of an orbit on a Poincar\'e surface of section $(z,\dot{z})$.

A particular example that we studied recently (Efthymiopoulos et al. 2015), is the magnetic bottle Hamiltonian

\begin{equation}
H=\frac{1}{2}(\dot{\rho}^{2}+\dot{z}^{2})+\frac{1}{2} \rho^{2}+ \frac{1}{2} \rho^2 z^2 - \frac{1}{8} \rho^4 + \frac{1}{8} \rho^2 z^4 - \frac{1}{16} \rho^4 z^2 + \frac{1}{128} \rho^6
\end{equation}

\noindent which has the special feature that the second frequency $\omega_{2}$ is zero.
This Hamiltonian has been used in explaining the orbits generating the aurora near the poles of the earth.
 The orbits in the plane $(\rho, z)$ are either ordered (Fig. 1a) or chaotic (Fig. 1b).

\begin{figure}
\centering
\includegraphics[width=\textwidth]{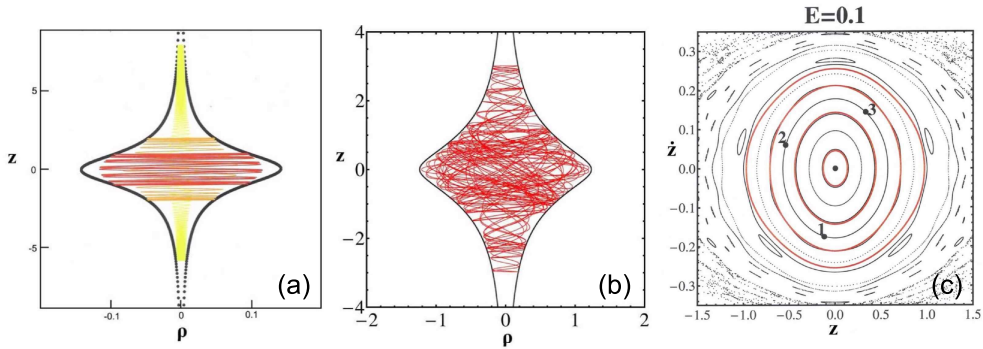}
\caption{Ordered and chaotic orbits in the magnetic bottle model. (a) 3 ordered orbits with different colours, (b) one chaotic orbit, (c) Invariant curves around the origin $O(0,0)$, numerical (black) and theoretical (red), for an energy $E=0.1$. An orbit starting on an invariant curve (1) has its images (2, 3,...) on the same invariant curve.}
\label{fig:01}
\end{figure}

On a surface of section $(z,\dot{z})$ the ordered orbits are represented by invariant curves around the origin (Fig. 1c), or around a set of islands of stability (Fig. 2a,b). Namely the successive intersections of the orbit by the surface of section lie on a given invariant curve (Fig. 1c) or on successive islands (Fig. 2a,b). The theoretical invariant curves (red) found by using a truncated integral $\Phi_{s}$ of order $s=12$ are very close to the numerical invariant curves (black) for small values of the energy (Fig. 1c). However for somewhat larger value of the energy there are islands of stability and the usual theoretical curves fail to represent these islands (Fig. 2a). The main new result of our recent paper (Efthymiopoulos et al. 2015) was that there exists a resonant form of the new integral $\Phi$ (in the case $\omega_{2}=0$) which represents very well the islands of a particular resonance (4/1 in the present case) and also the nonresonant invariant curves around the origin (Fig. 2b).

\begin{figure}
\centering
\includegraphics[width=\textwidth]{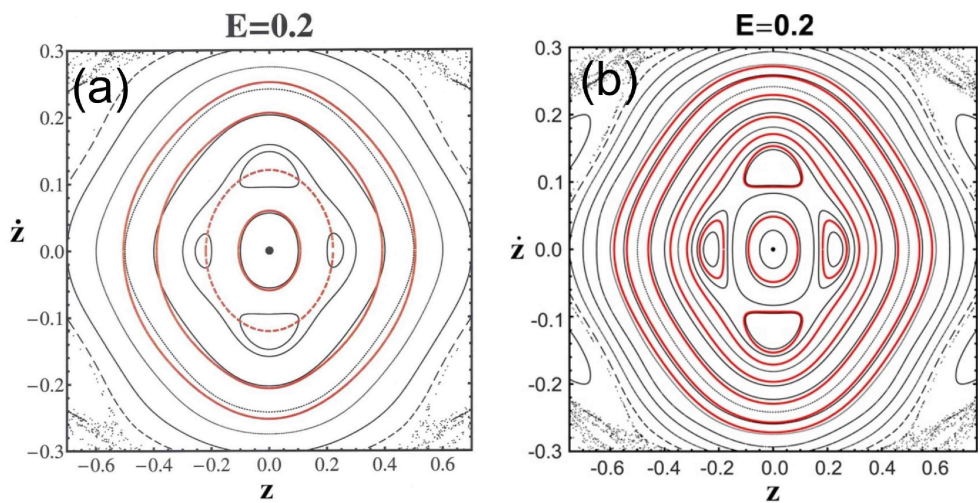}
\caption{Invariant curves for $E=0.2$, numerical (black) and
theoretical (red) for the Hamiltonian (6). (a) The nonresonant form
of the integral $\Phi$ represents well the invariant curves around
the origin, but not the islands. (b) The resonant form of the
integral $\Phi$ represents both the
 invariant curves around the origin and the 4 islands.}
\label{fig:02}
\end{figure}

On the other hand for larger values of the energy the origin is unstable and around it there is an important chaotic region (Fig. 3a) (The intervals of the energy where the origin is unstable are given in the Appendix). Orbits in this chaotic region are chaotic and their successive intersections with the surface of section seem to be random.

\begin{figure}
\centering
\includegraphics[width=\textwidth]{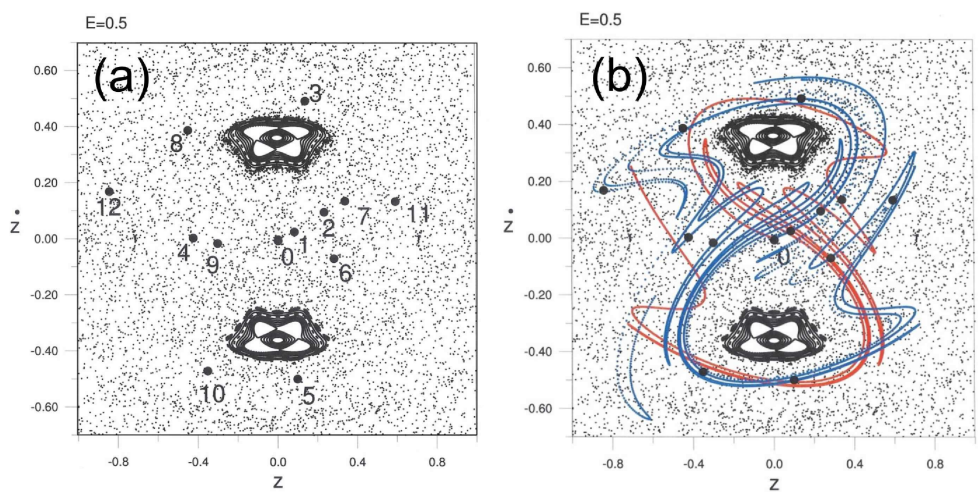}
\caption{When the energy is E=0.5 the central equilibrium $O(0,0)$ is unstable. (a) Then the images of nearby initial conditions (point 1) seem to be scattered in a random way. (b) However as the point (1) is on the unstable asymptotic curve from $O$, all its images are on the same asymptotic curve (blue). (For comparison we give also the stable asymptotic curve (red))}
\label{fig:03}
\end{figure}

However, around the unstable point at the origin, the new integral $\Phi$ is accurate (convergent) and not only formal. In fact, in the unstable case one frequency in the second order Hamiltonian (2) is real and the other is imaginary, thus no small divisors appear. This case was indicated already by Cherry (1926) and it was studied in detail by Moser (1956, 1958) and Giorgilli (2001).

In particular the asymptotic curves from the central unstable point  (blue and red curves in Fig. 3b) are given by convergent series. Although these curves are complicated, there are given theoretically with arbitrary accuracy. Thus, an orbit with initial conditions on an asymptotic curve has all its intersections on the same asymptotic curve.
 In this way the successive points 1,2,3,\dots etc of the orbit of Fig. 3a lie exactly on the unstable asymptotic curve (blue) emanating from the unstable periodic orbit $O$ (Fig. 3b).

The problem that was not solved in the previous papers was to find the limits of the convergence of the series $\Phi$. This subject was studied by us (Efthymiopoulos et al. 2014, Harsoula et al. 2015, Contopoulos and Harsoula 2015) in recent years and we describe its main points in sections 2 and 3. Then in section 4 we describe briefly some more recent results, namely an application to the chaotic spiral arms emanating  from the Lagrangian points $L_{1}$, $L_{2}$ of barred spiral galaxies.


\section{Moser invariant curves and chaos in mappings}
A simple map that has an unstable point at the origin is the hyperbolic H\'enon map (da Silva Ritter et al. 1987):

\begin{eqnarray}
\nonumber x' &= \cosh (\kappa) x + \sinh (\kappa) (y-\frac{x^2}{\sqrt{2}})     \\
y' &= \sinh (\kappa) x + \cosh (\kappa) (y-\frac{x^2}{\sqrt{2}})
\end{eqnarray}

When the parameter $\kappa$ is equal to $\kappa=1.43$ the eigenvalues of the origin $(x=y=0)$ are
$\lambda_{1}=e^{\kappa}=4.1787$ and $\lambda_2 = 1 / \lambda_{1}=0.2393$. A normal form series for the map (7) can be computed as follows: We change the variables to:

\begin{equation}
u= (x+y) / \sqrt{2} \hspace{2mm}, \hspace{5mm} v=  (x-y) / \sqrt{2}
\end{equation}
and introduce a near identity transformation

\begin{eqnarray}
u & = & \Phi_{1} (\xi,\eta)= \xi + \Phi_{1,2}(\xi,\eta) + \Phi_{1,3}(\xi,\eta) + \dots  \nonumber  \\
v & = & \Phi_{2} (\xi,\eta)= \eta + \Phi_{2,2}(\xi,\eta) + \Phi_{2,3}(\xi,\eta) + \dots
\end{eqnarray}

\noindent (where $\Phi_{i,s}$ $(i=1, 2)$ contain the terms of $s$ degree) such that the new variables $(\xi, \eta)$ are transformed linearly

\begin{equation}
\xi ' = \Lambda(c) \xi \hspace{2mm}, \hspace{5mm} \eta ' = \frac{1}{\Lambda(c)} \eta
\end{equation}

with

\begin{equation}
\Lambda  =  \lambda_{1} + w_{2} c + w_{3} c^2 + ...  \hspace{2mm}, \hspace{5mm} \frac{1}{\Lambda}  =  \lambda_{2} + v_{2} c + v_{3} c^2 + ...
\end{equation}

where

\begin{equation}
c = \xi \eta = constant
\end{equation}

\noindent and $w_{s}$, $v_{s}$ are constants independent of $c$.

Thus in the new variables ($\xi,\eta$) the hyperbolae (12) are invariant curves (Fig. 4a). The formulae  that give the Moser transformations (9)-(11) are given by da Silva Ritter et al. (1987).

\begin{figure}
\centering
\includegraphics[width=\textwidth]{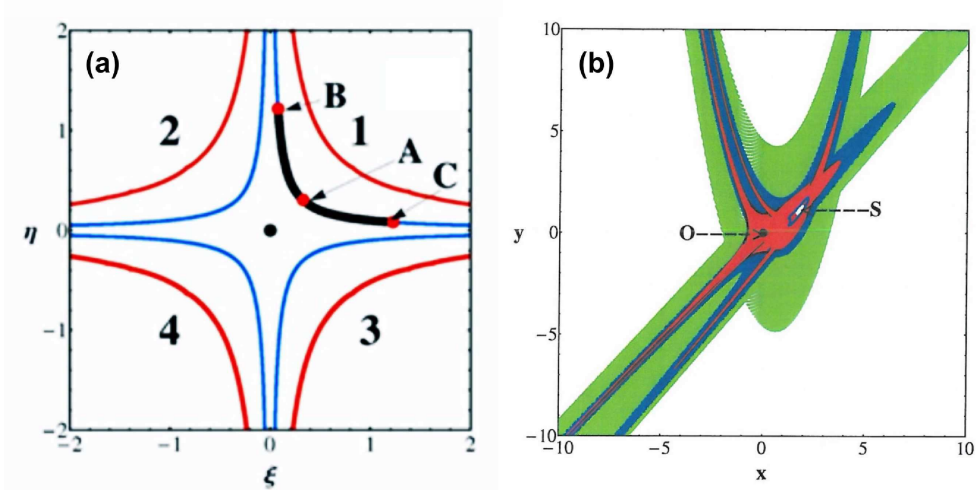}
\caption{(a) The invariant curves in the plane $(\xi, \eta)$ are hyperbolae $c = \xi \eta$ in the sectors 1, 2, 3 and 4. A point on a hyperbola has its images and pre-images on the same hyperbola. B is the first image of A and C its first pre-image. (b) The Moser domain of convergence (red) in the variables $(x,y)$ for $\kappa=1.43$. $S$ represents a stable periodic orbit. This is surrounded by a last KAM curve (black) and inside it is the inner limit of the Moser domain. All orbits with initial conditions outside the red region (in the intervals $-10 \leq x \leq 10, -10 \leq y \leq 10$) have their first images (green) and their second images (blue) approaching the boundaries of the domain of convergence.}
\label{fig:04}
\end{figure}

We found the domain of convergence of the series (9)-(11) by using
the d' Alembert criterion (Efthymiopoulos et al. 2014). Namely we
calculated the limits of the ratios

\begin{equation}
\frac{|\Phi_{i,s}|}{|\Phi_{i,s+1}|}
\end{equation}

\noindent along a given direction $\phi$, where $\xi=R \cos\phi$ and $\eta = R \sin \phi$, in the $(\xi, \eta)$ plane.

We found numerically that the limits of the ratio (13) depend on $c$ but not on $\phi$. For $\kappa=1.43$ the series converge for $|c| \leq 0.49$. Thus the convergence domain in the plane $(\xi, \eta)$ is limited by the hyperbolae $c=+0.49$ (regions 1 and 4 in Fig. 4a) and $c=-0.49$ (regions 2 and 3). Any orbit with initial conditions inside this domain, has its images and pre-images along a hyperbola passing through this point (Fig. 4a).

If we back-transform the hyperbolae of Fig. 4a to the original variables $(x,y)$, using Eqs. (8)-(9), we find the image of the domain of convergence of the Moser series in the plane $(x,y)$ (red in Fig. 4b). This domain is limited outwards by the images of the curves $c=\pm 0.49$ of the regions 2,3,4  of Fig. 4a. However there is also an inner limit, which is the image of the curve $c=0.49$ of the region 1, in the ($\xi,\eta$) plane. Thus there is an inner region (white in Fig. 4b) where the series do not converge. This region is around a stable periodic point $S$.

In Fig. 5 we have drawn a number of invariant curves $\xi \eta=c$ as they are mapped in the variables $(x,y)$. We call these curves ``Moser invariant curves''. In particular the curves $c=0$ (red) represent the axes $\eta=0$ and $\xi=0$. These curves correspond  to the stable and unstable asymptotic curves of the point $(0,0)$ and  extend to infinity within the domain of convergence, because  the series (9) (and of course also (10)-(11)) converge all the way to infinity if $\eta=0$ or $\xi=0$. The images of the curves $\xi=0$ and $\eta=0$ intersect each other at an unlimited number of homoclinic points that can be found analytically.

\begin{figure}
\centering
\includegraphics[width=.5\textwidth]{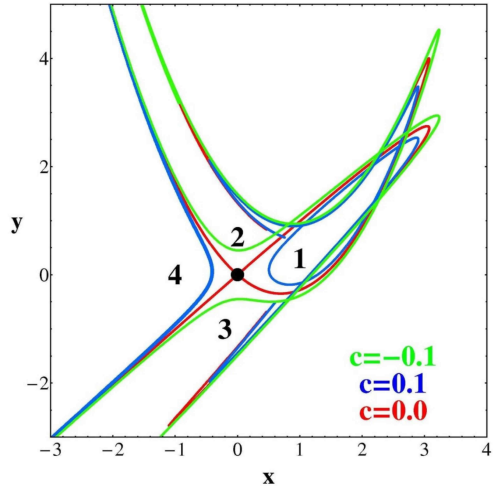}
\caption{A number of Moser invariant curves on the plane $(x,y)$, that are images of hyperbolae of Fig. 4a in the sectors 1, 2, 3 and 4.}
\label{fig:05}
\end{figure}

Any point in the domain of convergence of Fig. 4b has its images inside the same domain. On the other hand orbits with initial conditions outside the domain of convergence have images that approach closer and closer the outer limits of this domain. This is seen in Fig. 4b where we have mapped a grid of initial conditions $(-10,10)  \times (-10,10)$ outside the convergence domain. Their first images are in green and their second images are in blue (covering also the inner green regions). Higher order images are congested even closer to the outer limits of the domain of convergence. Thus the outer limits of the domain of convergence act as an attractor for the orbits outside this domain.

The orbits close to the stable invariant point $S$ are either ordered, forming invariant curves (KAM curves), or chaotic (around unstable periodic points of higher order). The KAM curves can be represented (approximately) by the usual series expansions around $S$ of the form of the ``third'' integral (Contopoulos 2002). The last KAM curve (black curve in Fig. 6a) is outside the inner limit of the domain of convergence (blue curve). Between this limit and the last KAM curve there are Moser invariant curves, that are completely inside the last KAM curve (Fig. 6a). However,  there are Moser curves that intersect the last KAM curve infinitely many times and extend very far from it (Fig. 6b). Finallly there are Moser invariant curves completely outside the last KAM curve (Fig. 6c).

\begin{figure}
\centering
\includegraphics[width=\textwidth]{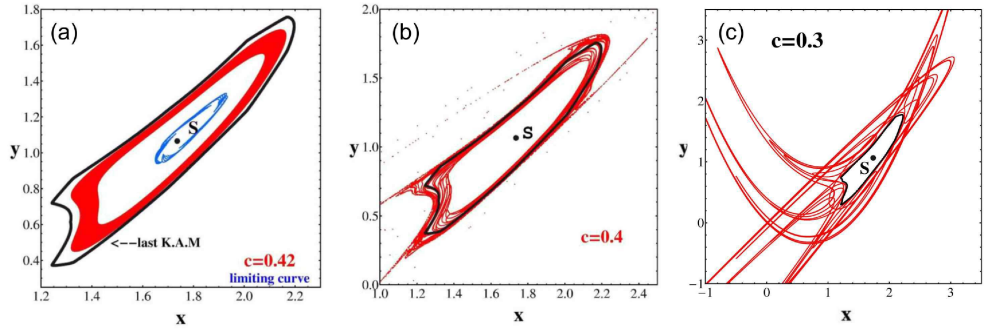}
\caption{(a) A Moser invariant curve (red) completely inside the last KAM curve (black) and outside the limiting curve (blue). (b) A Moser invariant curve intersecting the last KAM curve, infinitely many times, but extending also to large distances. (c) A Moser invariant curve completely outside the last KAM curve.}
\label{fig:06}
\end{figure}

In conclusion the Moser invariant curves are of three types (a) orbits completely inside the last KAM curve $(0.42 \lesssim c \lesssim 0.49)$, (b) orbits intersecting the last KAM curve $(0.32 \lesssim c \lesssim 0.42)$ and (c) orbits completely outside the last KAM curve $(c \lesssim 0.32)$,  It is of interest to note that the first type of Moser invariant curves contains both ordered orbits (and these orbits can be represented both by Moser series and KAM series), and chaotic orbits (that cannot be represented by KAM series). We emphasize that there is no contradiction in the fact that the Moser invariant curves are able to represent both ordered and chaotic orbits. For details see Harsoula et al. (2015). On the other hand inside the boundary ($c \geq 0.49$)  i.e. closer to the point $S$ the Moser series around the origin $O$ do not converge, while there are KAM invariant curves that are represented (approximately) by the third integral type of series (but there are also small domains of chaotic orbits around high order unstable periodic orbits that cannot be represented by such series).

The main application of the Moser series regards the orbits starting close to the origin $O$ where chaos is dominant. In this case the successive images of an initial condition close to $O$ look quite random (Fig. 7a). However, all these points lie on a particular Moser invariant curve curve (Fig. 7b) that can be given accurately analytically. Thus these orbits can be given by analytical formulae. The only indication of chaos is that although all the iterates lie on the same Moser curve, the distance between successive iterates increases at every iteration. Furthemore, the Moser invariant curves make many oscillations that extend to large distances, therefore some points on them may be at considerable distances from the center.

\begin{figure}
\centering
\includegraphics[width=\textwidth]{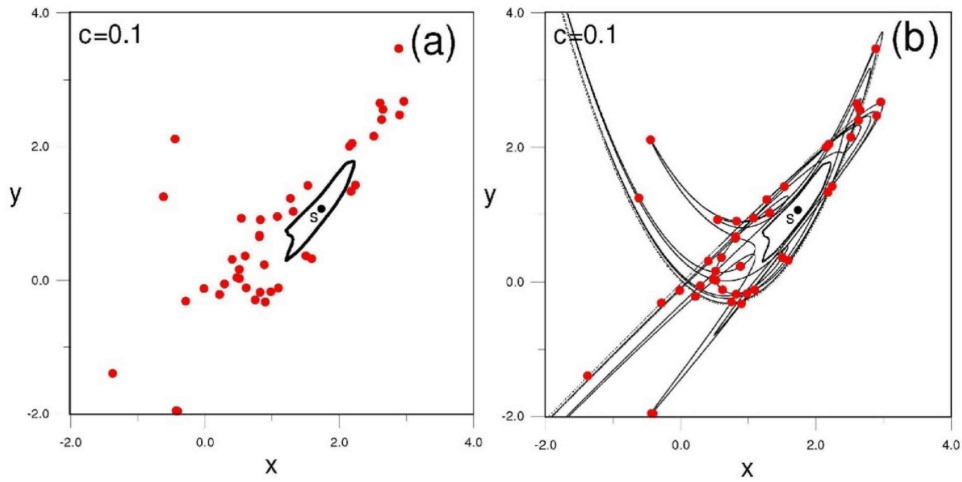}
\caption{(a) The images of a point close to the origin $O$ are
distributed in a random way. (b) However all these images lie on a
particular Moser invariant curve.} \label{fig:07}
\end{figure}

In this connection we must point out that the truncation error of the series is not uniform along the invariant Moser curves. In fact, at a fixed truncation order $r$, the error is small for the part of the hyperbola near the diagonal $\xi=\eta$, but it increases exponentially for the parts of the hyperbola approaching asymptotically the axes. As a consequence, we find that in order to accurately represent a segment of the hyperbola of length $s$ around the diagonal using the Moser series we need to reach a truncation order $r$ which increases exponentially with $s$. This property is a manifestation of the exponential divergence of nearby chaotic orbits. Namely, in order to numerically integrate accurately a chaotic orbit up to a fixed time $t$, we need to know the initial conditions with a number of digits growing exponentially with $t$. On the other hand, in order to obtain analytically with a given precision a chaotic orbit up to a time $t$, we need to specify a number of coefficients in the series growing exponentially with $t$. Thus the chaotic character of the orbits manifests itself in the necessity to have either very accurate initial conditions (for a numerical calculation), or a large number of terms of the Moser series (for an analytical calculation).

A particular type of orbits along Moser invariant curves are the periodic orbits. In fact near the homoclinic points there is an infinity of periodic orbits. Such orbits were found by da Silva Ritter et al. (1987) in the following way. The homoclinic point $H$ (Fig. 8a) has an infinity of images along the stable manifold of the periodic orbit $O(c=0)$. Invariant curves with $c$ close to zero intersect themselves at points near the homoclinic point $H$. The images of such a particular intersection (let us call it point 0) are the points 1, 2, ... If we join these points with the point $S$ we form successive angles that have an average value (rotation angle) $\phi$. If $\phi={2\pi}/{n}$ the $n^{th}$ image coincides with 0 and we have a periodic orbit of period $n$.


\begin{figure}
\centering
\includegraphics[width=\textwidth]{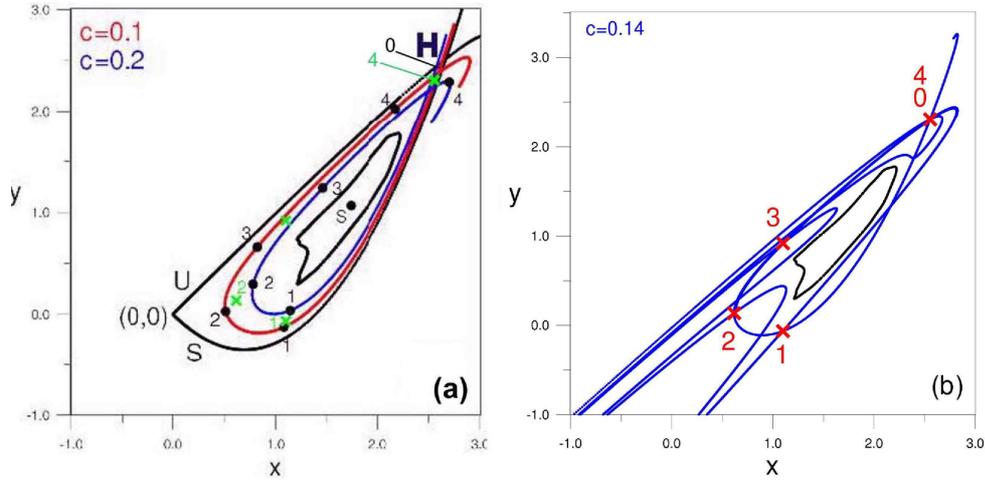}
\caption{(a) A method to find a periodic orbit of period 4 that has a point close to the homoclinic point $H$ is the following: We use two invariant curves ($c=0.1$ (red) and $c=0.2$ (blue)) that intersect themselves close to $H$ (point 0) and have images 1,2,3,4 where the point 4 is on the left of 0 for $c=0.1$ and on the right of the corresponding intersection 0 for $c=0.2$. Thus between them there is a periodic orbit $1, 2, 3, 4 \equiv 0$ of period 4 (green points). (b) An invariant curve (with $c \simeq  0.14$)
  that passes an infinite number of times from the 4 points of the periodic orbit of period 4.}
\label{fig:08}
\end{figure}

In Fig. 8a we plot two invariant curves with $c=1$ (red) and $c=0.2$ (blue). The first curve has a rotation angle $\phi <{2\pi}/{4}$ (the point 4 is on the left of the original point $O$) and the second curve has a rotation angle $\phi > {2\pi}/{4}$ (its point 4 is on the right of the corresponding point $O$). Thus between these two curves there exists a curve with rotation number exactly $\phi={2\pi}/4$ and this is found by interpolation. The green points $1,2,3,4 \equiv 0$ form a periodic orbit of period 4 with $c \simeq 0.14$. Closer to the homoclinic point $H$ there are periodic orbits of period 5,6,... up to infinity. In fact the homoclinic points themselves can be considered as belonging to a periodic orbit with period $n=\infty$.

The invariant curve that passes through the four points of the periodic orbit with rotation number $\phi={2\pi}/4$, has many intersections with itself, but it passes an infinite number of times from the points $1,2,3,4 \equiv 0$ (Fig. 8b). All the periodic orbits close to the homoclinic point $H$ are unstable. These orbits were generated from the stable point $S$ for smaller values of the parameter $\kappa$.

In Fig. 9 we give the characteristics of the various families 6,5,4,3,2 together with the characteristics of the periodic orbit $S$ and of the homoclinic point $H$. The two families of period 4 are generated from the stable family $S$ at $\kappa=1.317$. The upper family is initially stable while the lower one is unstable.  The stable family becomes unstable for $\kappa=1.375$ at a period doubling bifurcation, and then all the families that were produced by a cascade of further  bifurcations (with intervals between successive bifurcations decreasing by the universal ratio $\delta=8.72$) become unstable beyond $\kappa=1.383$. The same happens with all the families of order higher than 4. All these families are congested close to the homoclinic point $H$.


\begin{figure}
\centering
\includegraphics[width=.5\textwidth]{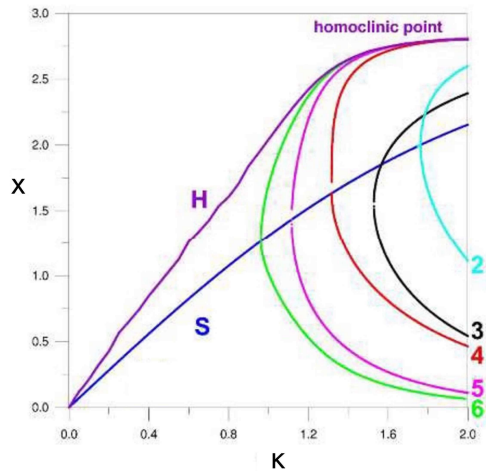}
\caption{Characteristics giving $x$ as functions of $\kappa$ for the families $S$, 6, 5, 4, 3 and 2, and for the homoclinic point $H$.}
\label{fig:09}
\end{figure}


Of special interest are the families of period 2 because when these
families are bifurcated  from the original family $S$ (for
$\kappa=1.76$), the orbit $S$ itself becomes unstable. The stable
family 2 becomes unstable at $\kappa=1.84$ and then follows a
cascade of periodic doubling bifurcations so that beyond
$\kappa=1.86$ all the bifurcated families become unstable.

When $S$ becomes unstable it has its own asymptotic curves and new Moser invariant curves close to them.
 We found a Moser transformation that gives these new asymptotic invariant curves (Contopoulos and Harsoula 2015). We found also a Moser domain of convergence of these transformations which is completely inside the Moser domain of convergence around the orginial point $O$ (Fig. 10a). In this case ($\kappa=2$) there is no inner limit of the domain of convergence around $O$.


\begin{figure}
\centering
\includegraphics[width=\textwidth]{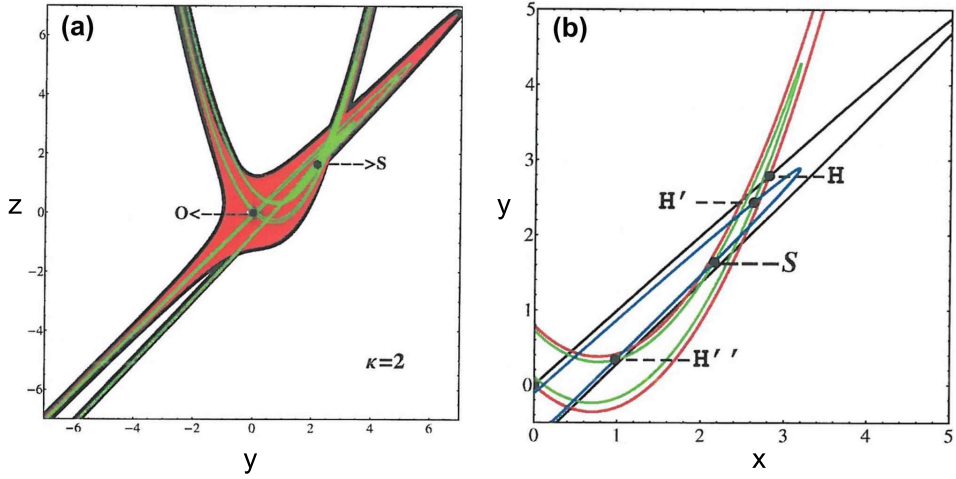}
\caption{(a) The Moser domains of convergence for orbits around $O$ (red) and around $S$ (green) for $\kappa=2$. The green domain is completely inside the red domain. (b) The asymptotic curves from $S$, for $\kappa = 2$ intersect at homoclinic points $H'$ and $H''$.}
\label{fig:10}
\end{figure}


For $\kappa=2$ the asymptotic curves from $S$ intersect themselves at the homoclinic points $H'$ and $H''$ (Fig. 10b). The eigenvalues of $S$ are $\lambda_{1}=-3.21$ and $\lambda_{2}=1/ \lambda_{1}=-0.31$, therefore this orbits is hyperbolic unstable. Then the successive points on the asymptotic curves are alternatively above and below $S$.

The asymptotic curves from the orbit $S$ intersect the asymptotic curves from the orbit $O$ at heteroclinic points. All the homoclinic and heteroclinic points are found analytically with a very good accuracy as compared with the numerical results (Contopoulos and Harsoula 2015).

The heteroclinic intersections do not appear immediately after the orbit $S$ becomes unstable. In fact when $S$ just becomes unstable there is still a last KAM curve around $S$, as it was when $S$ was stable, and closed invariant curves inside it (but at a certain distance from $S$). Then the asymptotic curves from $S$ remain for ever inside the last KAM curve. This situation occurs for a small interval of values of $\kappa$  $(1.76 \leq \kappa \leq1.79)$. But when $\kappa$ exceeds $\kappa=1.79$ the last KAM curve around $S$ is destroyed and the asymptotic curves from $S$ intersect with the asymptotic curves from $O$.

Moser domains of convergence appear around every unstable periodic orbit. E.g. such domains of convergence appear around the unstable periodic orbit 2 that bifurcates from $S$ and also when the stable orbit 2 becomes unstable. Thus we can imagine that the red Moser domain of convergence of Figs. 4b and 10a is full of smaller Moser domains of convergence around the various unstable orbits in the original domain. The corresponding orbits can be given analytically by more than one Moser series. This has been checked in cases where we have intersections of Moser invariant curves around $O$ with curves around $S$ (Contopoulos and Harsoula 2015).

Our conclusion is that inside the Moser domains, where most orbits are chaotic we can find analytic expressions for such orbits and describe in detail their chaotic behavior.


\section{Hamiltonian Systems}

While in simple mappings, like the hyperbolic H\'enon map, the
convergence along the asymptotic curves of a hyperbolic orbit goes
to infinity, in Hamiltonian systems the convergence extends only up
to a finite distance. Thus, it is not certain that one can find
analytically the homoclinic intersections of the asymptotic curves.
In fact although Vieira  and Ozorio de Almeida (1996) and Bongini et
al (2001) applied the method of Moser in hamiltonian cases, they
could not find even the first homoclinic point. They only approached
it when the Moser series was truncated at high orders. However by
applying a method of analytic continuation we could extend the
applicability of the Moser method and calculate theoretically
several homoclinic points (Efthymiopoulos et al. 2014).

As an example we used the Hamiltonian

\begin{equation}
H = \frac{1}{2} p^2 - {\omega_{0}}^2 [1+\epsilon (1+p) \cos{\omega t}] \cos{\psi}
\end{equation}

\noindent which represents a perturbed pendulum. For $\epsilon = 0$ we have the Hamiltonian of a pendulum

\begin{equation}
H = \frac{1}{2} p^2 - {\omega_{0}}^2 \cos{\psi}
\end{equation}

If we use a dummy action $I$, conjugate to the angle $\phi = \omega t$ we can write an equivalent Hamiltonian

\begin{equation}
H = \frac{1}{2} p^2 + \omega I -  {\omega_{0}}^2 [1+\epsilon(1+p) \cos{\phi}] \cos{\psi}
\end{equation}

For $\epsilon = \omega = 1$, ${\omega_{0}}^2=0.08$ a Poincar\'e surface of section is given in Fig. 11a by taking $\phi = 2 \pi n$, ($n$ = 0, 1, 2,...). Most of the central region of Fig. 11a is chaotic, but there are also some islands of stability near the axis $\psi = 0$ and also regular orbits above and below the chaotic domain. The chaotic behaviour is around the asymptotic curves from the unstable periodic orbit $P(mod 2\pi)$ ($P$ at $\psi_{0} = - \pi$ or $P'$ at $\psi_{0} = \pi$). These asymptotic curves intersect at an infinity of homoclinic points above and below the axis $PP'$. (The stable and unstable asymptotic curves undergo large oscillations in the upper part of Fig. 11b, but smaller oscillations in the lower part.)


\begin{figure}
\centering
\includegraphics[width=\textwidth]{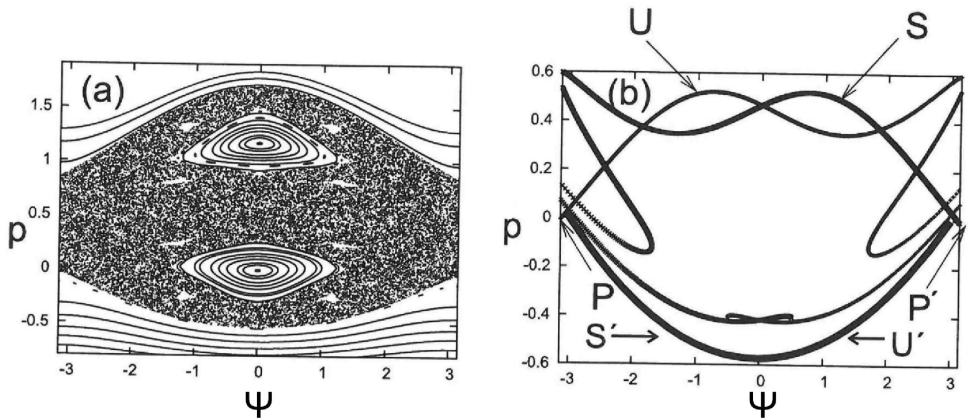}
\caption{(a) The surface of section $(\Psi, P)$ for orbits in the Hamiltonian (14) for $\epsilon = \omega = 1$ and ${\omega_{0}}^2 = 0.08$. Most orbits in a domain around $p=0$ are chaotic. (b) The asymptotic curves from the periodic orbit ($P$ at $\psi_0 = - \pi$) (which is the same as the orbit $P'$ at $\psi_0 = \pi$.)}
\label{fig:11}
\end{figure}


We measure now the angles $u$ from the point $\psi_{0} = \pi$, i.e. we have $\psi = u - \pi$. If we develop $\cos{\psi}$ in powers of $u$ we find

\begin{equation}
H = \frac{1}{2} p^2 + \omega I  + 0.08 [1+\epsilon (1+p) \cos{\phi}] [1-\frac{u^2}{2}+\frac{u^2}{24}-\dots]
\end{equation}

The lowest order term of this Hamiltonian is

\begin{equation}
H_{0} = \frac{1}{2} (p^2 - {\omega_{0}}^2 u^2)
\end{equation}

\noindent and if we use diagonal variables $(\xi,\eta)$ through

\begin{equation}
p = \frac{\sqrt{\omega_{0}}  (\xi +\eta)}{\sqrt{2}}  \; \; \;  u = \frac{(\xi -\eta)}{\sqrt{2 \omega_{0}}}
\end{equation}

we find

\begin{equation}
H_{0} = \omega_{0} \xi \eta
\end{equation}

Thus the Hamiltonian (15) becomes

\begin{equation}
H = \omega I  + \omega_{0} \xi \eta + H_{1}
\end{equation}

\noindent where $H_{1}$ is the perturbation $H_{1}=H_{1}(\phi, I, \xi, \eta)$  (In this particular case $H_{1}$ does not depend on $I$).

Then we find new variables $(\phi', I', \xi', \eta')$ such that

\begin{equation}
H = \omega I'  + {\omega_{0}}' \xi' \eta' + Z(I', \xi'  \eta')
\end{equation}

\noindent i.e the perturbation depends only on $I'$ and on product $c=\xi'  \eta'$ and does not depend on $\phi'$.

Then the quantities $I'$ and $c=\xi'  \eta'$ are integrals of motion. The transformations

\begin{equation}
(\phi, I, \xi, \eta) = \Phi (\phi', I', \xi', \eta')
\end{equation}

\noindent are given by a normalization method in the form of series (Efthymiopoulos et al. 2014).

In these variables the successive iterates in the plane $(\xi', \eta')$ are along hyperbolae. In particular the asymptotic curves are the axes $\eta' = 0$ and $\xi' = 0$ (Fig. 12). On the other hand in the original variables $(\xi, \eta)$ the asymptotic curves are curved and intersect at homoclinic points like $H_{0}$. If now we come back to the original variables $(\psi,p)$ on the surface of section we find curves that are similar to the curves $U$ and $S$ of Fig. 11b close to the origin (Fig. 13),
  but further away they deviate considerably. If we increase the order of the normalizing series (22) we approach closer to the homoclinic point $H_{0}$ but do not reach it.
  This is due to the fact that the convergence of the series is limited up to two points $A$ and $A'$ along the asymptotic curves.


\begin{figure}
\centering
\includegraphics[width=.5\textwidth]{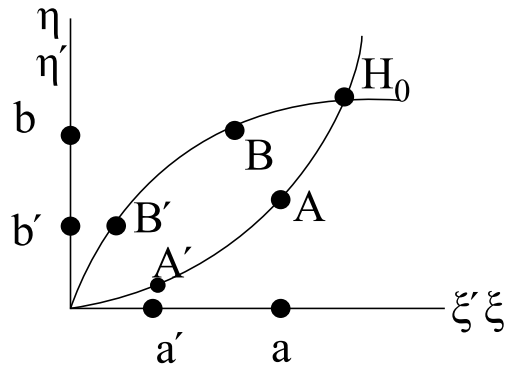}
\caption{The asymptotic curves in the variables $(\xi', \eta')$ are the axes $\xi '= 0$ and $\eta'=0$, while in the variables $(\xi , \eta)$ they are curves intersecting at the homoclinic point $H_0$ (schematically). A point $\alpha$ in on the $\xi'$ axis beyond the convergence domain has an $m-th$ preimage $\alpha'$ inside the convergence domain, whose image on the $(\xi, \eta)$ plane is the point $A'$. Then the $m-th$ image of $A'$ is $A$. Thus $A$ can be found by analytic formulae as the image of $\alpha$. Similary $B$ is the image of $b$.}
\label{fig:12}
\end{figure}



\begin{figure}
\centering
\includegraphics[width=.5\textwidth]{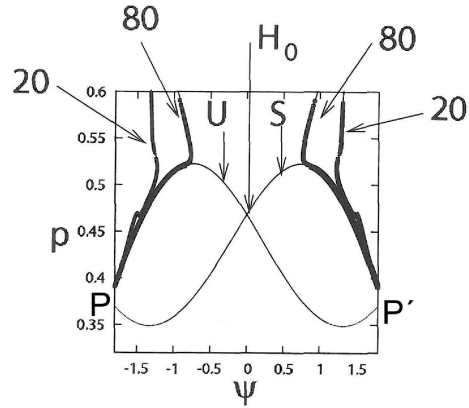}
\caption{The Moser series for the invariant curves from the point $P$ ($U$ = unstable) and $P'$ ($S$=stable) truncated at orders 20 and 80 are close to the numerical asymptotic curves up to some distance but then they deviate considerably and do not approach very close the homoclinic point $H_0$.}
\label{fig:13}
\end{figure}



\begin{figure}
\centering
\includegraphics[width=.8\textwidth]{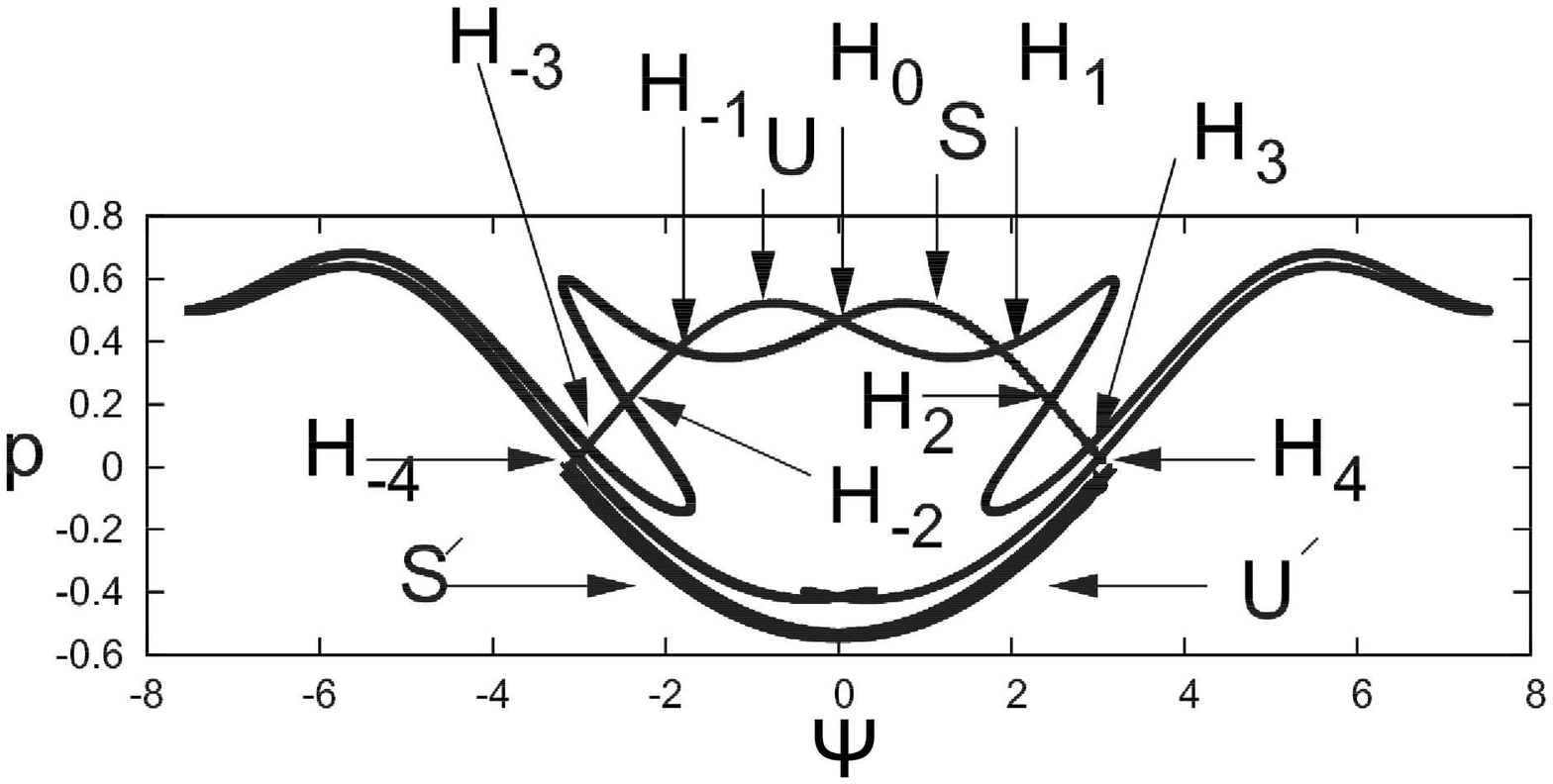}
\caption{Our new method allows the calculation of the Moser invariant curves up to a large extent, allowing the analytical calculation of the homoclinic points $H_0, H_1, H_2, H_3, H_4, H_{-1}, H_{-2}, H_{-3}, H_{-4}$.}
\label{fig:14}
\end{figure}


In fact the series giving the variables $(\phi, I, \xi, \eta)$ in terms of $(\phi ', I', \xi ', \eta ')$ converge in the complex plane of $\Psi$ up to a limit $|Im (\Psi)| \leq \sigma$, $\sigma > 0$ while the real values $Re(\Psi)$ vary from 0 to $2\pi$.

In order to overcome this difficulty, which is due to the limits of the convergence, we use the following method (Efthymiopoulos et al. 2014). If we have a point $a$ on the $\xi$-axis of Fig. 12, where the Moser formulae do not converge, we find a pre-image $a'$ close to the origin (say the m-th preimage),
 where the series converge and we find the corresponding point $a'$ on the plane $(\xi, \eta)$. Then by proceeding in steps along which there is no convergence problem we proceed from $a'$ to its m-th image, which is the required image $A$ of $a$. In the same way we proceed from a point $b$ along the axis $\eta$ to find its image $B$. The images and the pre-images can be found by a series based on the Lie operator defined by the Hamiltonian (15). Thus, the whole method corresponds to a method of analytic continuation that allows us to go beyond the limits of convergence of the original series due to singularities in the complex domain.

Using this method in the problem of the perturbed pendulum with $m=3$ and 4 steps (4 successive transformations) we could find in Fig. 14 accurately the asymptotic curves that were computed numerically in Fig. 11b up to at least 9 homoclinic points $(H_{0}, H_{1}, H_{2}, H_{3}, H_{4}, H_{-1}, H_{-2}, H_{-3}, H_{-4})$.

The conclusion from the above studies is that in principle we can find analytically the asymptotic curves (and nearby invariant curves) up to an arbitrarily large length. Thus we can find analytically the chaotic orbits for an arbitrarily long time. However, in practice we have computational limitations related to the growing complexity of the series.


\section{Application to barred spiral galaxies}

A particular application of this new method has been made recently (Harsoula et al. 2015), in order to connect the Moser domain of convergence around the Lagrangian points $L_1$ and $L_2$ of barred-spiral galaxies with the chaotic spiral arms beyond them. A particular N-body simulation of a barred-spiral galaxy  produced by Voglis et al. (2006) (named model QR2), shows a concentration of stars along spiral arms emanating from the ends of the bar (Fig.15a). The orbits of the stars supporting the spiral arms are chaotic. A particular chaotic orbit starting close to $L_1$ is superimposed in Fig.15a. This orbit stays for a long time (compared to a Hubble time), close to the spiral arms and the outer parts of the bar, before escaping from the system. In general the stars stay longer close to the apocentra (or pericentra) of their orbits. Therefore the apocentra (or pericentra) define the spiral arms (Fig. 15b) which are density waves, i.e. the stars of the spiral arms are continuously replaced by other stars.


\begin{figure}
\centering
\includegraphics[width=\textwidth]{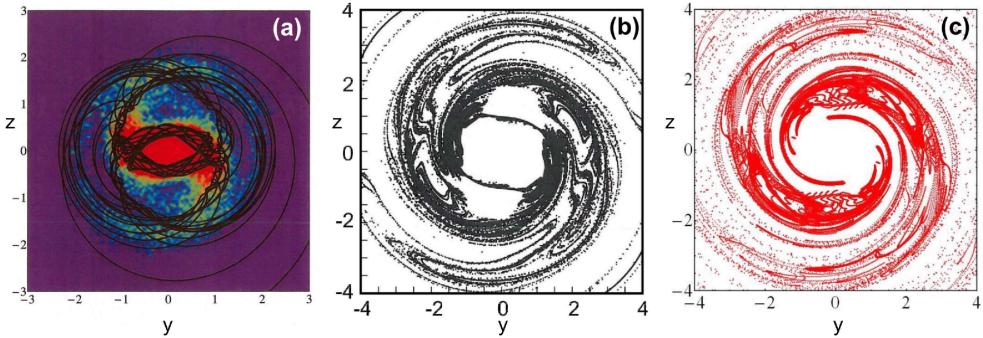}
\caption{(a) The N-body surface density of a N-body simulation of a barred galaxy (colours) shows the bar and the spiral arms emanating from the ends of the bar. Superimposed is an orbit starting close to the Lagrangian point $L_1$. (b) The apocentric manifolds from the ends of the bar define the basic structure of the spiral arms. (c) The spiral arms generated by the theoretical asymptotic curves and their neighbourhoods are quite close to the numerical asymptotic curves of Fig. 15b.}
\label{fig:15}
\end{figure}


We now apply the method described in the previous sections to find the analytical asymptotic curves emanating from the Lagrangian points $L_{1}$ and $L_{2}$ at the end of the bar (Fig. 15c). These analytical asymptotic curves and the Moser invariant curves close to them make many oscillations, back and forth, but their overall form is very similar to the distribution of the apocentra (and pericentra) of the orbits shown in Fig. 15b. The details of these calculations will be given in a future paper.

A simple theoretical model of the spiral arms is provided by finding an averaged Hamiltonian and a corresponding mapping, using a method introduced by Hadjidemetriou (1991, 2008).  We start with a Hamiltonian representing the N-body, distribution on the plane of rotation, of the form:

\begin{equation}
H = \frac{{p_r}^2}{2} + \frac{{p_\phi}^2}{2 r^2} - \Omega_{p} p_{\phi} + \Phi(r,\phi)
\end{equation}

\noindent where $r$ and $\phi$ are polar coordinates, in a frame rotating with pattern velocity $\Omega_P$ and $p_r$, $p_\phi$
are the corresponding momenta. The potential $\Phi$ consists of an axisymmetric part $\Phi_0$ and a mode $m=2$ in the form

\begin{equation}
\Phi = \Phi_0 (r) + \Phi_1 (r) \cos{2\phi} + \Phi_2 (\phi) \sin{2\phi}
\end{equation}

Then we develop the Hamiltonian around the corotation radius $r_0$
setting $r=r_0 + \delta r$, $p_\phi = p_0 + J_\phi$, and introduce a
pair of action angle variables $(J_r,\phi_r)$ by the
transformation

\begin{equation}
\delta r = \sqrt{2 J_r / \kappa_r} \sin(\phi_r),  p_r = \sqrt{2 \kappa_r J_r} \cos(\phi_r)
\end{equation}

\noindent where $\kappa_r$ is the epicyclic frequency given by the
formula:

\begin{equation}
\kappa_r = \sqrt{{\partial \Phi_{eff}^2}/ {\partial r^2}}
\end{equation}

\noindent with an effective axisymmetric component

\begin{equation}
\Phi_{eff} = \frac{{p_0}^2}{2 r^2}+\Phi_0 (r)
\end{equation}

Thus we derive a Hamiltonian $H=H(J_{\phi}, \phi, J_r , \phi_r )$. Then we average this Hamiltonian over the angles $\phi_r$, while $J_r$ is a constant. The orbits in this averaged Hamiltonian are found by solving the equations of motion.

We then use the Hadjidemetriou method (1991,2008),  which consists of finding a 2D-mapping that has the same fixed points and the same stability indices with the given Hamiltonian. In our particular case of a galaxy, we find an approximate mapping for this system, which is the standard map

\begin{equation}
x_{1}' = x_{1}+x_{2}'  \hspace{2mm},\hspace{5mm} x_{2}' = x_{2}+ K \sin(x_{1})
\end{equation}

\noindent (without the modulo $2 \pi$). In the case of the model QR2 of Voglis et al. (2006), we find that the value of $K$ is $K = 2.68488$.

The origin of this standard map corresponds to the Lagrangian point $L_1$ (or $L_2$) of the spiral Hamiltonian.
 This point is an unstable periodic orbit, and we can now apply the Moser theory around it. Namely we find a transformation to new variables ($\xi, \eta$) in which the mapping is

\begin{equation}
\xi' = \Lambda(c) \xi \hspace{2mm},\hspace{5mm}  \eta' = \frac{1}{\Lambda(c)} \eta
\end{equation}

\noindent where

\begin{equation}
\xi' \eta' = \xi \eta = c (constant)
\end{equation}

\noindent i.e. the mapping is along hyperbolae in the variables $(\xi, \eta)$. The formulae giving the variables $(x,y)$ are convergent whenever $|\xi \eta|=|c|$ is smaller than a critical value $c_{crit} \backsimeq 4$. The image of the domain of convergence $|\xi \eta| \leq c_{crit}$ in the plane $(x_1,x_2)$ is called now a ``Moser domain''. The most important result is the following: While in the variables $(x_1,x_2)$ of the standard map the Moser domain is given as a black region in Fig. 16a, in the original variables $(y, z)$ of the configuration space of the galactic model, the Moser domain has a spiral form (Fig. 16b).


\begin{figure}
\centering
\includegraphics[width=.8\textwidth]{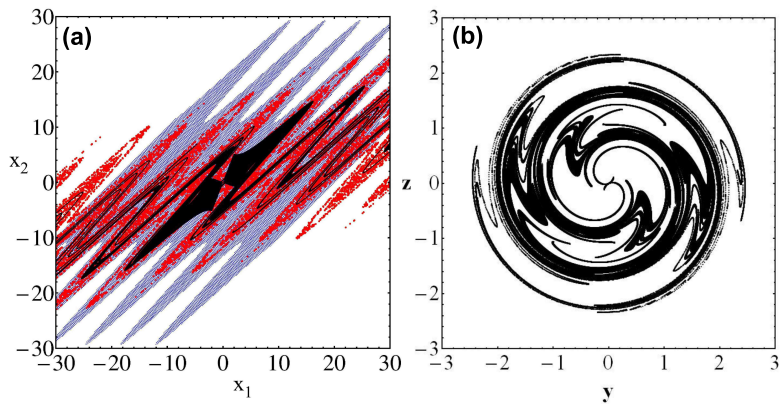}
\caption{(a) The Moser domain in the variables $(x,y)$ (black), and the first and third iterates of a grid of initial conditions ($-20 < x_1 < 20$ , $-20 < x_2 < -20$) (blue and red respectively.) (b) The Moser domain of convergence in the cofiguration space of the galaxy.}
\label{fig:16}
\end{figure}


It is of great interest that the theoretical form of the spirals, as given by the Moser domain, is very close to the observed spirals of the N-body simulation. The details of these calculations are given in Harsoula et al. (2016).

Finally, we take initial conditions outside the Moser domain we find that their iterates approach the outer limit of the Moser domain. This can be seen in Fig. 16a where we have taken a dense grid of
initial conditions with $-20 < x_1 < 20, -20 < x_2 < 20$ that fill an area larger than the limits of the N-body system. The first iterates of these initial conditions (blue) approach the Moser domain and the third iterates (red) approach it even closer (Fig. 16a). Higher order iterates approach very close the boundary of the Moser domain. Thus the boundary of the Moser domain acts as an attractor for the orbits that start outside this domain. On the other hand orbits starting inside the Moser domain cannot ever go outside it.

The fact that the boundary of the Moser domain acts as an attractor was found first in the case of a different mapping (the hyperbolic H\'enon map) by Contopoulos and Harsoula (2015). Thus, it seems that this phenomenon is quite general.

On the other hand we can see in Fig. 16a that the Moser domain extends to larger values of $x_1$ and $x_2$
 as the number of iterations increases. In this way the orbits can go to large distances and eventually they escape to infinity.
 However the orbits remain close to the spiral arms of the galaxy for a long time before escaping to infinity.
 This means that the chaotic orbits outside corotation remain sticky close and along the asymptotic manifolds for a long time before allowing their orbits to escape.

Our conclusion is that the asymptotic curves (manifolds) from the unstable periodic orbits (and the nearby invariant curves) that are given by analytical formulae can have important applications in particular problems of interest.


\appendix
\section{Appendix: Intervals of instability of the central periodic orbit}

The central periodic orbit $z=0$ of the magnetic bottle given by the Hamiltonian (6) is stable for $0<E<0.366882$. Beyond this energy, this periodic orbit becomes unstable for an interval $\Delta E_1$
and then undergoes transitions from stability to instability and vice versa as the energy approaches a critical value $E_{crit}= 0.592593$ at which the period tends to infinity. In fact the period of the orbit $z=0$ is found from the equation

\begin{equation}
\frac{1}{2} {\dot{\rho}}^2 + V(\rho,0)=E
\end{equation}

\noindent where

\begin{equation}
V(\rho,0)=\frac{1}{2} \rho^2 - \frac{1}{8} \rho^4 + \frac{1}{128} \rho^6
\end{equation}

The function $V(\rho,0)$ (Fig. 17a) has a maximum $E=E_{crit}=0.592593$ for $\rho_{max}=\rho_{crit}=1.692993$. The period is twice the time to go from $\rho_{min} (= - \rho_{max})$ to $\rho_{max}$, where $\rho_{max}$ is the root of the equation $V(\rho,0)=E$, i.e.

\begin{equation}
T = 2  \int_{\rho_{min}}^{\rho_{max}}\frac{d \rho}{\sqrt{2(E-V(\rho,0))}}
\end{equation}

When $E$ decreases and tends to zero the period decreases and tends to $T=2 \pi$ and when $E \rightarrow E_{crit}$ it tends to infinity (Fig. 17b). For $E>E_{crit}$ the period decreases and tends to zero when the energy tends to infinity (Fig. 17b).


\begin{figure}
\centering
\includegraphics[width=.9\textwidth]{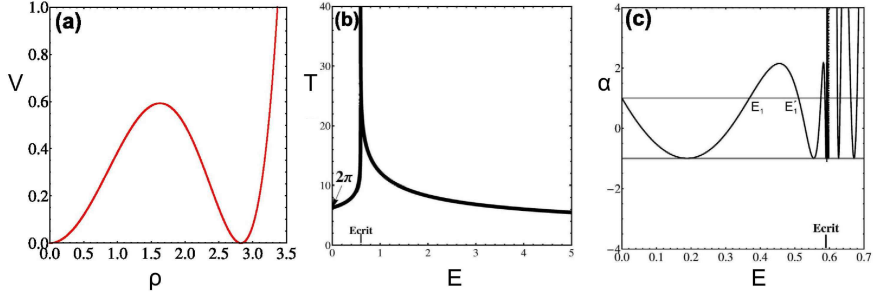}
\caption{(a) The function $V(\rho, 0)$. (b) The period $T$ of the orbits passing through the origin $(\rho = \Delta)$ as a function of the energy $E$. (c) The H\'enon stability index for the orbits passing through the origin $(\rho=0)$ as a function of the energy $E$.}
\label{fig:17}
\end{figure}


In order to find the stability of the periodic orbit $z=0$ we calculate the H\'enon stability index $\alpha$.
The orbit is stable if $-1<\alpha<1$ and unstable if $a>1$ (or $a<-1)$ (Fig. 17c). We see that as $E$ approaches $E_{crit}$ we have an infinity of intervals $(E_n , {E_n}')$ inside which the orbit is unstable (Table I).

\vspace{0.3cm}
\begin{tabular}{|c|c|c|c|c|c|c|}\hline
\multicolumn{4}{|l|}{\bfseries Table I} \\ \hline
n & $E_n$ & $\Delta E_n=|E_n-E_{crit}|$ & $\delta$=$\frac{\Delta E_n} {\Delta E_{n+1}}$ \\
\hline 1 &0.366882 & 0.225711   &13.03\\
\hline 2 &0.5752725 & 0.017320   &15.073\\ \hline  3&0.5914435 &  0.00114959 &  15.181 \\
\hline 4&0.5925169 & 0.0000756926  &15.176\\ \hline 5&0.592587605 & 4.98759 x $10^-6$ & 15.20\\
\hline 6&0.592592264575 & 3.28017 x $10^-7$  &15.188\\ \hline 7&0.592592570995 & 2.1597 x $10^-8$ & - \\
\hline
\end{tabular}\\
\vspace{0.3cm}

In particular the value $E=0.5$ of Fig. 3a is in the interval $(E_1 , {E_1}')$ where the orbit $z=0$ is unstable.

The phenomenon of infinitely many transitions to instability and stability was first observed by Churchill et al (1980) and by Contopoulos and Zikides (1983). Then Heggie (1983) has shown that the distances $\Delta E_n$ of the successive energies $E_n$ where we have transitions to instability from the critical energy $E_{crit}$ decrease by a factor

\begin{equation}
\delta = \frac{|E_n - E_{crit}|}{|E_{n+1} - E_{crit}|}
\end{equation}

\noindent which tends to a limiting number

\begin{equation}
\delta = exp \left[ \pi \sqrt{ \Big| {\frac{V_{\rho\rho}}{V_{zz}}}} \Big| \right]_{\rho=\rho_{max}, z=0}
\end{equation}

\noindent where $V_{pp}$ and $V_{zz}$ are the second derivatives of $V$ with respect to $\rho$ and $z$.

In the present case the theoretical value of Eq. (36) is $\delta=15.19$. If we compare this value with the numerical ratios of Table I we find that indeed the values of $\delta$ tend to this theoretical value as the order $n$ increases.

On the other hand for large values of $E$ the periodic orbit $z=0$ is unstable. As $E$ decreases this orbit becomes stable for the first time at $E=4.098$. As $E$ decreases further and tends to $E_{crit}$ we have an infinity of transitions from instability to stability (Fig. 17c). However in this case the ratio $\delta$ does not tend to $\delta=15.19$ as in the case of energies $E<E_{crit}$ for orbits around the origin.

\end{document}